\documentclass{article} 
\usepackage{iclr2026_conference,times}

\usepackage{amsmath,amsfonts,bm}

\def\eqref#1{equation~\ref{#1}}

\def\1{\bm{1}}

\DeclareMathAlphabet{\mathsfit}{\encodingdefault}{\sfdefault}{m}{sl}
\SetMathAlphabet{\mathsfit}{bold}{\encodingdefault}{\sfdefault}{bx}{n}

\usepackage{hyperref}
\usepackage{url}
\usepackage{multirow}
\usepackage{booktabs}
\usepackage{graphicx}
\usepackage{tabularx}
\usepackage{makecell}

\title{The 99\% Success Paradox: When Near-Perfect Retrieval Equals Random Selection}

\author{Vyzantinos Repantis, Harshvardhan Singh, Tony Joseph, Cien Zhang, Akash Vishwakarma,\\
\textbf{Svetlana Karslioglu, Michael Wyatt Thot, Ameya Gawde} \\
Meta Platforms Inc.\\
1 Hacker Way, Menlo Park, CA 94025, USA \\
\texttt{\{vrepantis,harshvardhan,tkjoseph,cienzhang,vishwaka,} \\
\texttt{  svekars,mthot,ameyagawde\}@meta.com}
}

\iclrfinalcopy
\begin{document}

\maketitle

\begin{abstract}
For most of the history of information retrieval (IR), search results were designed for human consumers who could scan, filter, and discard irrelevant information on their own. This shaped retrieval systems to optimize for finding and ranking more relevant documents, but not keeping results clean and minimal, as the human was the final filter. However, LLMs have changed that by lacking this filtering ability.

To address this, we introduce Bits-over-Random (BoR), a chance-corrected measure of retrieval selectivity that reveals when high success rates mask random-level performance. We measure selectivity as $BoR = \log_{2}\left(\frac{\mathrm{P}_{obs}}{\mathrm{P}_{rand}}\right)$, where $\mathrm{P}_{rand}$ is the hypergeometric baseline for the chosen success rule (here, coverage: $ \geq1 $ relevant in top-$K$).

On the 20 Newsgroups dataset, BM25 and SPLADE both report $>99$\% success at $K=100$ (coverage), yet $BoR \approx 0$, indicating random-level selectivity at that depth. When the expected coverage ratio $\left(\frac{K \cdot \bar{R}_{q}}{N}\right)$ exceeds 3--5, the baseline dominates and selectivity collapses. Downstream retrieval-augmented generation (RAG) evaluation confirms this pattern: LLM accuracy can degrade substantially at $K=100$, consistent with the near-zero BoR ceiling. In contrast, BoR remains positive on BEIR/SciFact and on MS MARCO (where 41 systems cluster within 0.2 bits of the theoretical ceiling despite a 13-point recall gap), confirming baseline predictions across sparse and large-scale settings. We further show that the collapse boundary applies to LLM agent tool selection, where small catalog sizes cause selectivity to vanish even with perfect selectors. These findings suggest reporting BoR alongside traditional metrics and reconsidering depth choices when additional retrieval provides negligible selectivity gains while inflating computational costs.
\end{abstract}

\section{Introduction}

Effective retrieval systems have played a pivotal role in accelerating the adoption of Large Language Models (LLMs) into widespread, specific use. By efficiently managing and accessing vast amounts of data, these systems enable LLMs to deliver precise and contextually relevant information, enhancing their utility across various domains, without the need for extensive fine-tuning. This synergy between retrieval systems and LLMs has facilitated their integration into industries such as finance, healthcare, and customer service, where accurate information is crucial.

Retrieval-augmented generation (RAG) and tool-using agents flip the assumptions of classical IR. The consumer is now an LLM, not a person, and the model does not skim. Every retrieved item it cannot ignore dilutes attention, costs tokens, and adds latency. The natural reflex is to enlarge the million-token context window and crank up $K$, but as we will show, more context is not always better: beyond a certain point, additional retrieved material actively degrades downstream output while still inflating cost.

This problem is most acute in \emph{LLM agent tool selection}, where the corpus is the catalog of tools, skills, or API endpoints available to the agent. Recent industry guidance reports tool definitions consuming many thousands of tokens before the agent has read a single user request. With small $N$ (50--500 tools) and several plausibly-relevant tools per task, the relevance density $\bar{R}_q/N$ is high---and the agent enters the regime our framework identifies as the \emph{collapse zone}, where even a perfect selector achieves near-zero selectivity over chance.

Retrieval systems can achieve near-perfect success rates while performing no better than random selection--a paradox often missed by traditional evaluation metrics. We introduce Bits-over-Random (BoR), a selectivity metric that reveals when retrieval systems perform no better than random selection. This has immediate implications for retrieval-augmented generation (RAG) systems.

As ~\cite{manning2008introduction} note, "recall is a non-decreasing function of the number of documents retrieved," yet the choice of retrieval depth $K$ is often an empirical application-dependent choice \cite{webber2010similarity}. In RAG pipelines, this matters: larger $K$ multiplies token costs while the chance baseline also strengthens with $K$. We show these forces can cancel, yielding near-perfect success with zero selectivity.

Consider a library of $N = 1{,}000$ books, with $R_q=10$ relevant to your query. Librarian A retrieves $K=20$ books, 6 of which are relevant (precision 30\%, recall 60\%, F1 40\%). Librarian B retrieves $K=12$ books, 4 of which are relevant (precision 33\%, recall 40\%, F1 36\%). Which librarian is more selective?

BoR measures selectivity in $bits$, as $BoR = \log_{2}\left(\frac{\mathrm{P}_{obs}}{\mathrm{P}_{rand}}\right)$: positive means better than random and 0 means random-level. Each bit represents a doubling of selectivity---a system with BoR=3 bits is $8\times$ more selective than random. Unlike precision@$K$ or NDCG, BoR is chance-corrected and doesn't mechanically increase with $K$ as it isolates selective value.

In our example, both librarians have a 100\% success rate under our chosen rule (coverage: $\geq 1$ relevant retrieved), so $\mathrm{P}_{obs}=1$ and $BoR = -\log_2 \mathrm{P}_{rand}$. BoR is 2.45 bits for Librarian A and 3.13 bits for Librarian B. Librarian B is more selective despite lower recall and F1. This better-than-chance effect is not visible to standard metrics and this tension is what we quantify.

\subsection{Contributions:}
\begin{enumerate}
    \item Demonstrate the 99\% paradox on 20 Newsgroups (20NG): $>99\%$ success@100 with $BoR \approx 0$ (random-level).
    \item Derive a depth identity that decomposes selectivity changes.
    \item Identify when selectivity collapses: when $\left(\frac{K \times \bar{R}_{q}}{N}\right)$ exceeds $3-5$, systems approach random-level performance.
    \item Validate the identity empirically: observed $\Delta BoR$ matches prediction within $<0.01$ bits across systems/datasets, showing architecture-agnostic behavior (BM25, SPLADE).
    \item Show that the collapse boundary applies to LLM agent tool selection, where small catalog sizes ($N = 50$--$500$) cause selectivity to vanish even with perfect selectors.
\end{enumerate}

\section{What Traditional Metrics Miss}
\label{traditional_metrics}

Recall rewards finding more relevant documents, but is blind to how many irrelevant items you had to pull into the context window to get them. Over-retrieval is actually rewarded. As \cite{manning2008introduction} note, "recall is a non-decreasing function of the number of documents retrieved". Yet the choice of retrieval depth K is often an empirical, application-dependent choice ~\cite{webber2010similarity}.

Precision measures the relevance of retrieved results and helps limit excessive retrieval. However, it fails to account for the inherent difficulty of the retrieval task. For instance, achieving a 10\% precision means something different if the corpus contains 10 relevant items out of 100 versus 10 relevant items out of $10,000$. Same precision, very different selectivity.

Ranking metrics (nDCG, RBP, MAP, ERR) penalize burying relevant items, but they do not penalize the presence of irrelevant items when the relevant item is also ranked highly. If you retrieve 100 items and the relevant one is at rank 1, nDCG can be perfect. Yet, RAG systems typically concatenate the top-K results into a single prompt. The LLM still has to read the other 99 items. Rankers optimize ordering, not volume. They don't reduce the token cost of stuffing K documents into the context.

In practice, teams end up juggling recall, precision, and ranking metrics. Each captures a different slice of behavior but none reflects the whole picture. There is no single framework that simultaneously accounts for how many items you retrieve, how big the corpus is, and how many items in the corpus are actually relevant to the query.

\section{Related Work}

Even though principled information-theoretic approaches for feature selection methods are available in Machine learning (~\cite{brown2012conditional, peng2005feature}), when it comes to finding the top-$K$ in information retrieval, the most common industry practice on involves tuning approaches like Grid Search Method and using precision and recall metrics to select the right $K$. Prior IR work notes recall's mechanical increase but lacks systematic calibration ~\cite{manning2008introduction}.

A further naive approach would be to substantially increase the value of $K$ and provide a large context window to latest generation LLM based AI models which supports really long context windows. ~\cite{liu2024lost} shows that LLM performance degrades when relevant information is positioned in the middle of long contexts. More retrieved documents don't always improve performance due to attention limitations. ~\cite{shi2023large} investigates the susceptibility of LLMs to irrelevant context in input prompts. The authors demonstrate that even state-of-the-art LLMs can be significantly influenced by unrelated or distracting information, leading to degraded performance on various tasks.

\section{Methodology}

Our framework relies on three fundamental definitions:
\begin{enumerate}
    \item \textbf{Coverage Rule:} A query succeeds at depth $K$ if the top-$K$ contains $\geq 1$ relevant item. Success is binary per query.
    \item \textbf{Success Rate:} Fraction of queries achieving success under the coverage rule. $\mathrm{P}_{\text{obs}}=1$ means perfect success under the chosen rule (finding $\geq 1$ relevant), \emph{not} retrieval of all relevant documents.
    \item \textbf{Hypergeometric Baseline:} Expected success rate for random document selection, accounting for query-specific relevance counts.
\end{enumerate}

\subsection{BoR and the Enrichment Factor}

Let $\mathrm{P}_{\text{obs}}$ be observed success and $\mathrm{P}_{\text{rand}}$ the random baseline. The Enrichment Factor ($EF$) is defined as $\mathrm{EF} = \mathrm{P}_{\text{obs}}/\mathrm{P}_{\text{rand}}$ measuring fold improvement over chance. EF is widely used in drug discovery to evaluate virtual screening~\cite{truchon2007evaluating}. BoR places EF on an additive information scale, where each additional bit doubles selectivity:

\begin{equation}
\mathrm{BoR} = \log_2(\mathrm{EF}) = \log_2\left(\frac{\mathrm{P}_{\text{obs}}}{\mathrm{P}_{\text{rand}}}\right)
\end{equation}

\subsection{Hypergeometric Baseline and Ceilings}

For coverage ($\geq 1$ relevant item retrieved) when query $q$ has $R_q$ relevant items, the exact random baseline is the probability that random selection retrieves at least one relevant document:

\begin{equation}
\mathrm{P}_{\text{rand}}(K; R_q) = 1 - \frac{\binom{N-R_q}{K}}{\binom{N}{K}}
\end{equation}

Macro-averaging yields the ceiling on achievable selectivity. When $\mathrm{P}_{\text{obs}} = 1$ (perfect success), $\text{BoR} = -\log_2(\bar{\mathrm{P}}_{\text{rand}}(K))$:

\begin{equation}
\text{BoR}_{\max}(K) = -\log_2(\bar{\mathrm{P}}_{\text{rand}}(K))
\end{equation}

where $\bar{\mathrm{P}}_{\text{rand}}(K) = \frac{1}{|Q|} \sum_q \mathrm{P}_{\text{rand}}(K; R_q)$ accounts for query-level variation in $R_q$.

When $R_q$ is unknown (common), we define the optimistic bound using $R_q=1$ for all queries:

\begin{equation}
\text{BoR}_{\text{opt}}(K) = \log_2\left(\frac{N}{K}\right)
\end{equation}

This optimistic ceiling provides an upper bound on achievable selectivity for any system operating on the same corpus with the same evaluation setup. Note that $\text{BoR}_{\max}$ uses actual $R_q$ values while $\text{BoR}_{\text{opt}}$ assumes $R_q=1$ throughout.

\subsection{Depth-Calibrated Identity}

This identity decomposes any BoR change into two competing forces: how much the success rate improves versus how much the random baseline becomes harder to beat as depth $K$ increases.

Within a fixed stratum (same evaluation setup: corpus size $N$, retrieval unit, and success rule), for any two depths $K_1$ and $K_2$:

\begin{equation}
\Delta\text{BoR} = \log_2\left(\frac{\mathrm{P}_2}{\mathrm{P}_1}\right) - \log_2\left(\frac{\bar{\mathrm{P}}_{\text{rand}}(K_2)}{\bar{\mathrm{P}}_{\text{rand}}(K_1)}\right)
\end{equation}

where $\mathrm{P}_1 = \mathrm{P}_{\text{obs}}(K_1)$ and $\mathrm{P}_2 = \mathrm{P}_{\text{obs}}(K_2)$ are the observed success rates at depths $K_1$ and $K_2$.

To make this identity tractable, we consider the rare-hit regime where relevance is sparse ($R_q \ll N$ and $K \ll N$). The exact hypergeometric baseline $\mathrm{P}_{\text{rand}}(K; R_q) = 1 - \binom{N-R_q}{K}/\binom{N}{K}$ accounts for drawing $K$ items without replacement. When the corpus is large relative to both retrieval depth and relevance count, each draw behaves nearly independently---removing a few documents doesn't meaningfully change subsequent draw probabilities.

Under this independence approximation, each of the $K$ draws has probability $R_q/N$ of selecting a relevant document, and probability $(1-R_q/N)$ of missing all relevant documents. Since the draws are independent, the probability that all $K$ draws miss the relevant documents is $(1-R_q/N)^K$. Therefore, the probability of hitting at least one relevant document becomes:

\begin{equation}
\mathrm{P}_{\text{rand}}(K; R_q) \approx 1 - (1-R_q/N)^K
\end{equation}

For small values, we apply standard approximations: $(1-x)^n \approx e^{-nx}$ and $e^{-y} \approx 1-y$. This yields $(1-R_q/N)^K \approx e^{-K \cdot R_q/N} \approx 1 - K \cdot R_q/N$ when $K \cdot R_q/N$ is small. Therefore:

\begin{equation}
\bar{\mathrm{P}}_{\text{rand}}(K) \approx K \cdot \bar{R}_q/N
\end{equation}

This simplifies our identity considerably:
\begin{equation}
\Delta\text{BoR} \approx \log_2\left(\frac{\mathrm{P}_2}{\mathrm{P}_1}\right) - \log_2\left(\frac{K_2}{K_1}\right)
\end{equation}

because $\bar{\mathrm{P}}_{\text{rand}}(K_2)/\bar{\mathrm{P}}_{\text{rand}}(K_1) \approx (K_2 \cdot \bar{R}_q/N)/(K_1 \cdot \bar{R}_q/N) = K_2/K_1$ in the rare-hit regime.

\textbf{Key Insight.} This identity reveals that selectivity changes decompose into two competing effects: system improvement versus random baseline strengthening. When success rates plateau ($\mathrm{P}_2 \approx \mathrm{P}_1$), the identity simplifies to $\Delta\text{BoR} \approx -\log_2(K_2/K_1)$. This (approximately) linear relationship makes the selectivity-depth trade-off transparent and predictable.

\textbf{Practical Implication.} Doubling depth ($K_2 = 2K_1$) costs approximately 1 bit of selectivity when success plateaus, since $-\log_2(2) = -1$. For example, a system achieving BoR=4 bits at $K=10$ will drop to BoR$\approx$3 bits at $K=20$ if success doesn't improve. Conversely, to \emph{maintain} selectivity when doubling depth, the success rate must also double ($\mathrm{P}_{\text{obs}}$ needs to increase by 100\%). Since success rates are bounded by 1, this becomes impossible once $\mathrm{P}_{\text{obs}} > 0.5$, explaining why selectivity inevitably degrades at larger depths.

\textbf{Extensions and Limitations.} The BoR framework extends to stricter success rules requiring at least $m$ relevant documents. In the rare-hit regime, the random baseline for the ``$\geq m$'' rule scales as $(K\,\bar{R}_q/N)^m/m!$ from Poisson approximation. When observed success plateaus ($P_2 \approx P_1$), doubling depth costs $m$ bits of selectivity:

\begin{equation}
\Delta\mathrm{BoR} \approx -m \log_2(K_2/K_1)
\end{equation}

We focus on $m=1$ because it matches common single-evidence retrieval scenarios and exposes the paradox most clearly. Outside the rare-hit regime (large $K \cdot \bar{R}_q/N$), the baseline saturates faster than linearly, and selectivity losses can exceed the $m$-bit approximation.

\subsection{Heuristic for Selectivity Collapse}

In the rare-hit regime, each draw has a small hit probability and $R_q/N \ll 1$. Under these conditions---many trials, small individual probability---the hypergeometric baseline is well-approximated by a Poisson model with rate $\lambda = K \cdot \bar{R}_q/N$, representing the expected number of relevant hits. For the coverage rule ($\geq 1$ relevant), the baseline becomes:

\begin{equation}
\bar{P}_{\mathrm{rand}}(K) \;\approx\; 1 - e^{-\lambda}
\quad\text{with}\quad
\lambda \;=\; K \cdot \frac{\bar{R}_q}{N}.
\end{equation}

This exponential form captures how random selection alone can make the task trivial as $K$ grows. To identify when selectivity becomes negligible, we solve for critical values of $\lambda$.

\textbf{Collapse boundary.} To find when selectivity collapses, we solve $\lambda = -\ln(1 - \bar{P}_{\text{rand}})$:

\begin{itemize}
\item \textbf{95\% random success:} $\bar{P}_{\text{rand}} = 0.95 \Rightarrow \lambda \approx 3.0$, \\
giving $\text{BoR}_{\max} = -\log_2(0.95) \approx 0.074$ bits;

\item \textbf{99\% random success:} $\bar{P}_{\text{rand}} = 0.99 \Rightarrow \lambda \approx 4.6$, \\
giving $\text{BoR}_{\max} = -\log_2(0.99) \approx 0.014$ bits.
\end{itemize}

\textbf{Practical rule.}
When $\lambda = K \cdot \bar{R}_q / N \in [3,5]$, random selection already succeeds $95$--$99\%$ of the time and the chance-corrected headroom is negligible (ceiling $< 0.1$ bits). In this \emph{collapse zone}, even a perfect system cannot show meaningful selectivity over chance.

\textbf{Scope.} This heuristic provides practical guidance for when selectivity collapses, using the Poisson/binomial approximation to the hypergeometric distribution. The $K \cdot \bar{R}_q/N \approx 3$--$5$ boundary offers operational thresholds rather than exact cutoffs. When $R_q$ is unknown, $\text{BoR}_{\text{opt}}$ (using $R_q=1$) provides an optimistic upper bound on achievable selectivity.

\section{Experiments and Results}

\subsection{Experimental Setup}

\textbf{Systems.} In our experimental setup, we compare BM25 (traditional lexical) with SPLADE~\cite{formal2022splade} (neural sparse retriever), using the public model \texttt{naver/splade-cocondenser-ensembledistil}. For SPLADE, we use document top-k=60, query top-k=60, max sequence length = 256, and batch sizes of 64 (documents) and 64 (queries). This pairing tests whether our findings hold across different retrieval architectures.

\textbf{Datasets.} We evaluate on three contrasting settings with different relevance densities. Here $\bar{R}_q$ denotes the mean number of relevant items per query, averaged across all queries in each dataset:
\begin{itemize}
\item \textbf{BEIR SciFact}~\cite{thakur2021beir} (N$\approx$5,185 documents): Low relevance multiplicity ($\bar{R}_q \approx 1.1$), representing typical sparse-relevance scenarios. Success is defined as top-$K$ containing at least one supporting evidence, counting SUPPORTS (2) and PARTIALLY SUPPORTS (1), excluding CONTRADICTS (0).

\item \textbf{MS MARCO Passage Ranking}~\cite{nguyen2016ms} ($N \approx 8.84$M passages): Large-scale web passage retrieval with extremely sparse relevance ($\bar{R}_q \approx 1$). We use Recall@1000 results reported in the literature for 41~systems spanning lexical baselines to state-of-the-art neural retrievers. This dataset tests whether BoR findings hold at production scale.

\item \textbf{20 Newsgroups}~\cite{lang1995newsweeder} (N=11,314 documents): High relevance multiplicity where queries match any document from their newsgroup category ($\bar{R}_q \approx 572$). We use class-based queries derived from document subject lines. Success is defined as top-$K$ containing at least one same-class document excluding the query itself. This creates conditions for potential selectivity collapse at K=100.
\end{itemize}

\textbf{Evaluation Protocol.} We measure success under the coverage rule ($\geq 1$ relevant retrieved) and compute BoR using exact hypergeometric baselines with per-query $R_q$ values. Dataset ceilings are macro-averaged: $\text{BoR}_{\max}(K) = \frac{1}{|Q|} \sum_q (-\log_2 P_{\text{rand}}(K; R_q))$. All results use seed=7 with 95\% confidence intervals from bootstrap resampling (n=5000).

\textbf{Downstream Check.} As a preliminary test of end-to-end impact, we evaluate a modern instruction-tuned LLM on retrieved contexts at K $\in$ \{10,100\} using temperature=0.0. We format queries as multiple-choice classification tasks and measure downstream accuracy on 20NG using 50 sampled queries per configuration.

\subsection{Experimental Results}

\textbf{Test 1: SciFact (The Benchmark Case).}
This is what most people expect: sparse relevance, the kind seen in real RAG systems. Both systems maintain strong selectivity even at $K=100$, with BoR staying above 5 bits. Predicted $\Delta$BoR values match observed changes to within $<0.01$ bits across all configurations. This confirms that when $\lambda = K \cdot \bar{R}_q / N \ll 1$ (well outside the collapse zone), retrieval systems can demonstrate meaningful selectivity over random chance.

\begin{figure}[ht]
\begin{center}
  \includegraphics[width=\columnwidth]{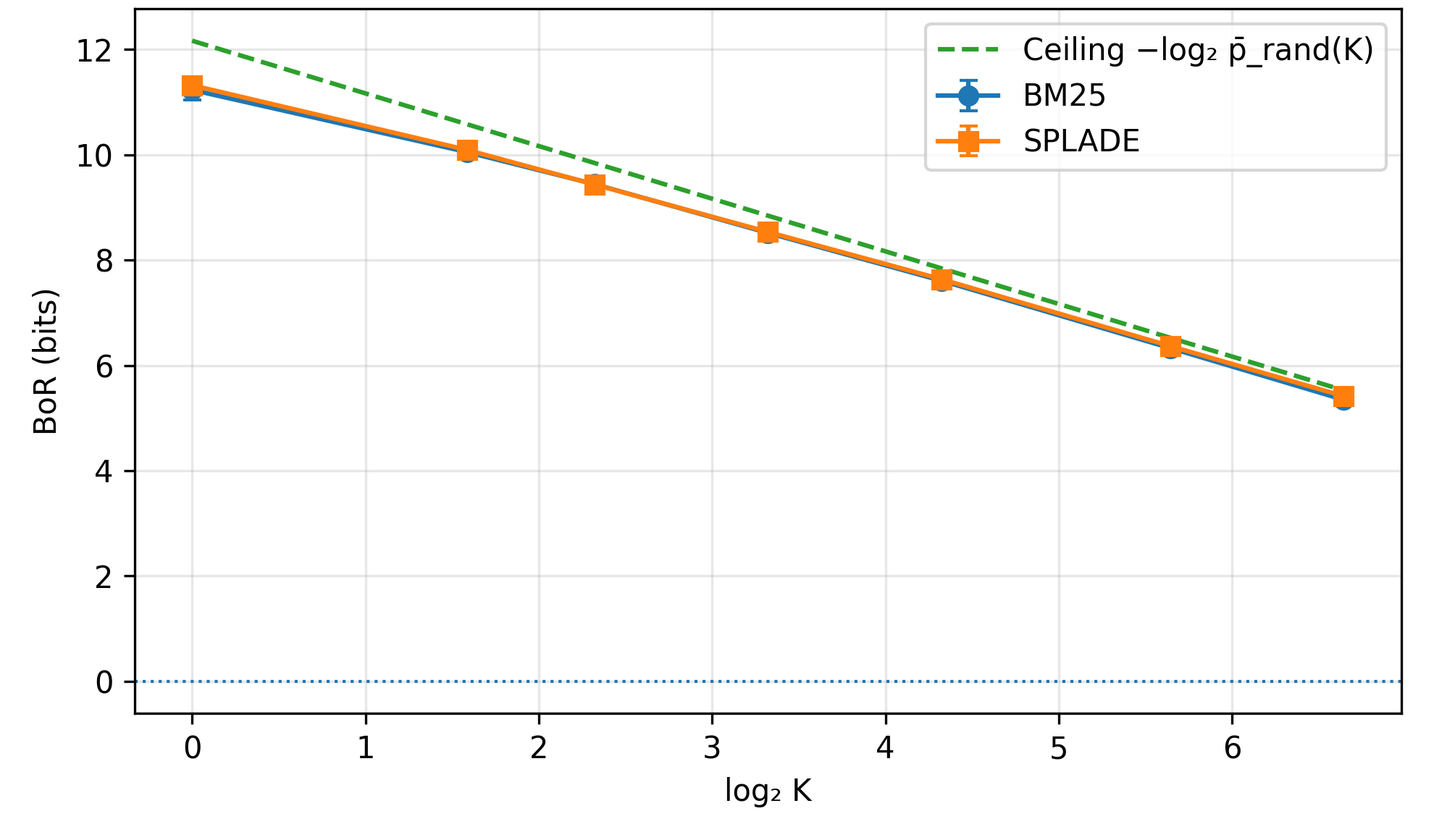}
\end{center}
\caption{BoR analysis on SciFact dataset shows sustained selectivity across retrieval depths. Both BM25 and SPLADE maintain high BoR values (5--11 bits), reflecting the dataset's sparse relevance structure.}
\label{fig:scifact_bor}
\end{figure}

But both BM25 and SPLADE operate very close to the theoretical ceiling. A 30-year-old algorithm nearly matches the modern neural system. Is SciFact just too easy? To investigate, we turn to a much larger benchmark.

\textbf{Test 2: MS MARCO (The Industrial Scale Test).}
With $N \approx 8.84$M passages and $\bar{R}_q \approx 1$, this is where large real-world systems operate. We computed BoR for 41~systems from the literature, from lexical baselines to state-of-the-art neural retrievers.

At $K=1000$, the theoretical ceiling is:
\begin{equation}
\text{BoR}_{\text{opt}} \approx \log_2\left(\frac{8.84 \times 10^6}{1000}\right) \approx 13.11 \text{ bits}
\end{equation}

All 41 systems cluster within 0.2 bits of this ceiling.

\begin{table}[ht]
\begin{center}
\renewcommand{\arraystretch}{1.3}
\begin{tabular}{@{}lcc@{}}
    \toprule
    System & Recall@1000 & BoR (bits) \\
    \midrule
    BM25 & 85.7\% & 12.89 \\
    SPLADE & 97.9\% & 13.08 \\
    ColBERTv2 & 98.5\% & 13.09 \\
    SimLM & 98.7\% & 13.09 \\
    \bottomrule
\end{tabular}
\end{center}
\caption{BoR analysis on MS MARCO Passage Ranking ($K=1000$, $\text{BoR}_{\text{opt}} \approx 13.11$ bits). Despite a 13-point recall gap between BM25 and SimLM, the BoR difference is only 0.20~bits. Systems examined include SimLM, AR2, uniCOIL, ColBERTv2, SPLADE (multiple versions), I3~Retriever, TCT-ColBERTv2, RoDR w/ ANCE, DPR-CLS, ColBERTer, ANCE, SLIM/SLIM++, and BM25.}
\label{tab:msmarco_results}
\end{table}

BM25 gets 85.7\% recall. SimLM (state-of-the-art) gets 98.7\% recall. That is a 13-point recall gap, but the BoR difference is only 0.20~bits. A three-decade-old lexical algorithm and cutting-edge neural systems are nearly indistinguishable in chance-corrected selectivity at this depth. This highlights diminishing returns from retriever improvements alone when relevance is sparse ($\lambda \approx 0.0001$, far from the collapse zone).

Both systems still show meaningful selectivity ($\text{BoR} > 12$~bits). To see what collapse looks like, we need an extreme test: a dataset where relevance is abundant, not rare.

\textbf{Test 3: 20 Newsgroups (The Stress Test).}
The 20 Newsgroups dataset has 20 topical categories. We set up an extreme scenario: treat all documents in the same category as ``relevant.'' With 11,314 documents split across 20 classes, that is about $\bar{R}_q \approx 572$ relevant documents per query (over 5\% of the corpus). At $K=100$:

$$\lambda = \frac{K \cdot \bar{R}_q}{N} = \frac{100 \times 572}{11{,}314} \approx 5.1$$

Random selection alone would succeed $\sim$99\% of the time.

\begin{table}[ht]
\begin{center}
\renewcommand{\arraystretch}{1.4}
\begin{tabularx}{\columnwidth}{@{}l l X X X X X X@{}}
    \toprule
    \makecell[l]{Dataset} & \makecell[l]{K} & \makecell[l]{BoR \\ Ceiling} & \makecell[l]{BM25 \\ Success} & \makecell[l]{BM25 \\ BoR} & \makecell[l]{SPLADE \\ Success} & \makecell[l]{SPLADE \\ BoR} & \makecell[l]{$\Delta$BoR \\ (10$\rightarrow$100)} \\
    \midrule
    \textbf{20NG} & 10 & 1.31 bits & 94\% & 1.22 & 95\% & 1.23 & $-1.22$ \\
    \textbf{20NG} & 100 & 0.01 bits & 100\% & 0.01 & 100\% & 0.01 & \multicolumn{1}{c}{---} \\
    \midrule
    \textit{SciFact} & 10 & 8.84 bits & 80\% & 8.52 & 81\% & 8.53 & $-3.12$ \\
    \textit{SciFact} & 100 & 5.52 bits & 89\% & 5.36 & 93\% & 5.41 & \multicolumn{1}{c}{---} \\
    \bottomrule
\end{tabularx}
\end{center}
\caption{BoR Analysis Results: System Performance Across Datasets and Depth. At $K=100$ on 20 Newsgroups, both systems achieve 100\% success but only 0.01 bits of selectivity---the ceiling has collapsed.}
\label{tab:bor_results}
\end{table}

The predicted $\Delta$BoR from theory matches reality within 0.01~bits. Perfect success rate, essentially zero selectivity.

\begin{figure}[ht]
\begin{center}
  \includegraphics[width=\columnwidth]{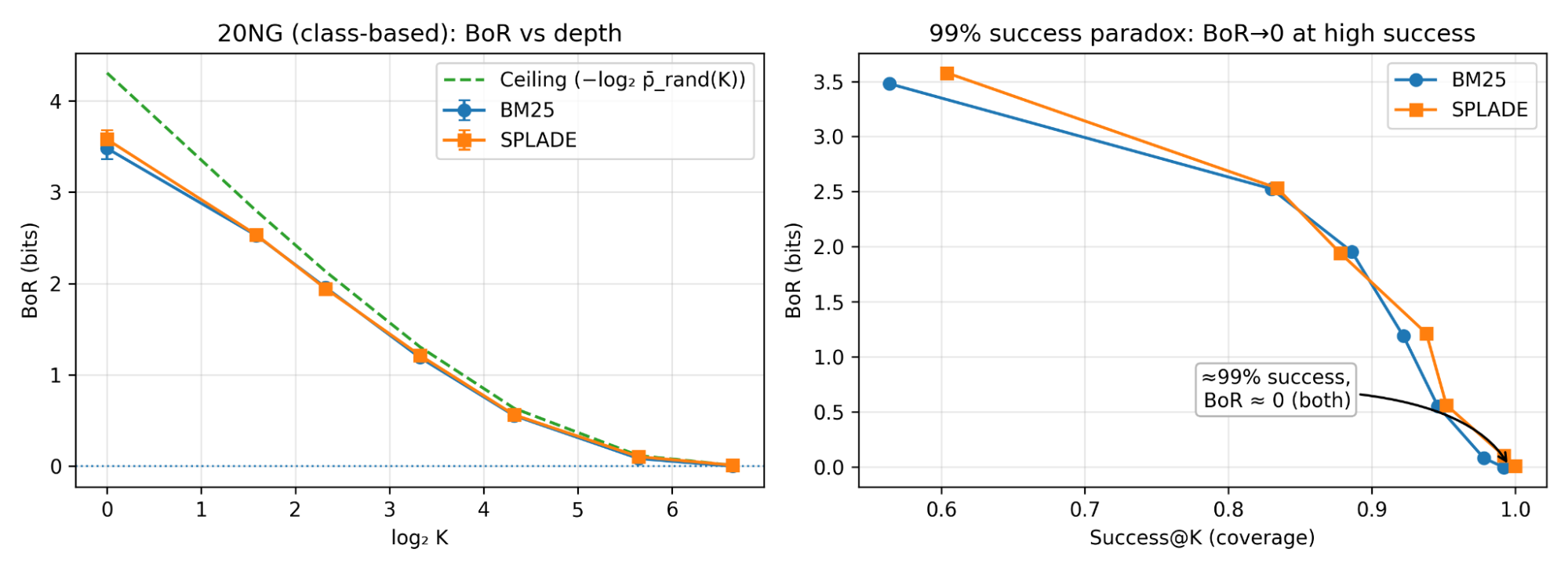}
\end{center}
  \caption{BoR analysis on 20 Newsgroups demonstrates the 99\% success paradox. Both BM25 and SPLADE show declining selectivity with increased depth (left) and convergence to random performance at high success rates (right).}
  \label{fig:image2}
\end{figure}

\textbf{When Perfect Success Fails: RAG Downstream Check.}
We tested this directly with a modern instruction-tuned LLM on the 20 Newsgroups collapsed scenario. Setup: multiple-choice classification task, 50 queries per configuration, temperature=0.0.

\begin{table}[ht]
\begin{center}
\renewcommand{\arraystretch}{1.3}
\begin{tabular}{@{}lcccc@{}}
    \toprule
    System & Accuracy (K=10) & Accuracy (K=100) & Success@K & Token Cost \\
    \midrule
    BM25 & 66\% & 50\% & 94\%$\rightarrow$100\% & 10$\times$ increase \\
    SPLADE & 68\% & 58\% & 95\%$\rightarrow$100\% & 10$\times$ increase \\
    \bottomrule
\end{tabular}
\end{center}
\caption{RAG downstream check on 20 Newsgroups. Despite success rates approaching 100\% at $K=100$, LLM accuracy drops by 10--16 percentage points while token costs increase 10$\times$.}
\label{tab:rag_downstream}
\end{table}

This is the failure mode BoR detects. The system pays 10$\times$ the tokens for random-level selectivity, and the LLM is drowning in noise. When selectivity collapses, high success rates become meaningless---or worse, misleading.

\section{LLM Agent Tool Selection}

The 20 Newsgroups stress test may appear artificial---who retrieves documents where 5\% of the corpus is relevant? However, this scenario maps directly to a problem LLM agents face daily: tool selection.

\subsection{When Agents Choose Tools}

Recent industry guidance~\cite{anthropic2025toolselection} reports that tool definitions can consume 50,000+ tokens before an agent reads a single user request. In one documented example, 58~tools consumed approximately 55K tokens; with integrations such as Jira, setups exceeding 134K tokens have been observed.

The retrieval framework applies directly to tool selection:

\begin{table}[ht]
\begin{center}
\renewcommand{\arraystretch}{1.3}
\begin{tabular}{@{}lll@{}}
    \toprule
    Parameter & Document Retrieval & Tool Selection \\
    \midrule
    $N$ & Corpus size (thousands to millions) & Available tools (50--500) \\
    $K$ & Documents shown to LLM & Tools shown to LLM \\
    $R_q$ & Relevant documents & Applicable tools for task \\
    \bottomrule
\end{tabular}
\end{center}
\caption{Mapping retrieval parameters to tool selection. The critical difference: $N$ is small for tools, so the collapse boundary is reached much faster.}
\label{tab:tool_mapping}
\end{table}

\subsection{The Tool Selection Collapse}

Consider the 58-tool example with 3--5 typically relevant tools per task ($R_q = 4$):

\begin{table}[ht]
\begin{center}
\renewcommand{\arraystretch}{1.3}
\begin{tabular}{@{}lccccc@{}}
    \toprule
    Configuration & $K$ & $R_q$ & $\lambda$ & $\text{BoR}_{\max}$ (Exact) & Regime \\
    \midrule
    Show 5 tools & 5 & 4 & 0.34 & $\sim$1.7 bits & Meaningful selectivity \\
    Show 20 tools & 20 & 4 & 1.38 & $\sim$0.28 bits & Degraded \\
    Show all 58 & 58 & 4 & 4.0 & $\sim$0 bits & Collapse \\
    \bottomrule
\end{tabular}
\end{center}
\caption{Tool selection collapse for a 58-tool system ($R_q = 4$). When all tools are shown ($\lambda \approx 4$), even a perfect selector achieves near-zero selectivity.}
\label{tab:tool_collapse}
\end{table}

When all tool definitions are introduced simultaneously into the model's context, the system operates at $\lambda \approx 4$, deep into the collapse zone. Even a perfect tool selector achieves only $\sim$0.02~bits of selectivity over random chance. As noted in~\cite{anthropic2025toolselection}, the most common failures are wrong tool selection and incorrect parameters, especially when tools have similar names---precisely the failure mode predicted by selectivity collapse.

\subsection{The Pattern Extends Beyond Tools}

The collapse boundary is a property of the selection problem itself. When $\lambda = K \cdot \bar{R}_q / N$ reaches 3--5, selectivity collapses regardless of what is being selected:

\begin{table}[ht]
\begin{center}
\renewcommand{\arraystretch}{1.3}
\begin{tabular}{@{}lccccl@{}}
    \toprule
    Scenario & $N$ & $R_q$ & $K$ & $\lambda$ & Regime \\
    \midrule
    RAG (typical) & 10,000 & 1--2 & 10 & $\sim$0.002 & Healthy \\
    Tool selection (filtered) & 20 & 3 & 5 & 0.75 & Healthy \\
    Tool selection (show all) & 20 & 3 & 20 & 3.0 & Collapse \\
    API endpoints (show half) & 100 & 8 & 50 & 4.0 & Collapse \\
    58-tool example & 58 & 4 & 58 & 4.0 & Collapse \\
    \bottomrule
\end{tabular}
\end{center}
\caption{Collapse boundary across selection scenarios. Agentic tool selection hits the collapse zone far more easily than document retrieval due to small $N$.}
\label{tab:collapse_scenarios}
\end{table}

This explains why agentic systems struggle with tool selection far more than RAG systems struggle with document retrieval. The mathematics is unforgiving when $N$ is small: with a corpus of millions, $\lambda$ remains small even at large $K$; with 50--500 tools, collapse is reached quickly.

\section{BoR with Alternative Success Rules}

This paper focuses on Success@K (coverage: $\geq 1$ relevant item retrieved), but the BoR framework generalizes to other success rules, including Recall@K.

\begin{table}[ht]
\begin{center}
\renewcommand{\arraystretch}{1.3}
\begin{tabular}{@{}p{2.2cm}p{3.0cm}p{2.8cm}@{}}
    \toprule
    Metric & What It Measures & Best For \\
    \midrule
    \textbf{Success@K} & Did you get $\geq 1$ relevant item? (Binary) & RAG/QA where one good context suffices \\
    \textbf{Recall@K} & What fraction of all relevant items did you get? (Graded) & Tasks needing comprehensive coverage \\
    \bottomrule
\end{tabular}
\end{center}
\caption{Success@K versus Recall@K as success rules for BoR.}
\label{tab:success_vs_recall}
\end{table}

\textbf{BoR for Recall@K.} Instead of measuring the probability of $\geq 1$ hit, we measure the expected fraction retrieved:

\begin{equation}
\text{BoR}_{\text{recall@K}} = \log_2\left(\frac{\text{observed\_recall@K}}{\text{expected\_recall@K\_random}}\right)
\end{equation}

For sparse relevance ($R_q \ll N$), the expected random recall is approximately $K/N$.

\textbf{Example.} A query has $R_q = 10$ relevant items in a corpus of $N = 1{,}000$ documents. A retriever finds 4 relevant items in the top-$K = 20$:
\begin{itemize}
    \item Observed recall $= 4/10 = 0.4$
    \item Random baseline $\approx K/N = 20/1{,}000 = 0.02$
    \item $\text{BoR}_{\text{recall@K}} = \log_2(0.4/0.02) = \log_2(20) \approx$ \textbf{4.32 bits}
\end{itemize}

The depth-calibrated identity (Equation~5) also extends to Recall@K with minor adjustments for the different success rule. We focus on Success@K in this paper because it matches the most common single-evidence RAG use case. Our MS MARCO analysis (Table~\ref{tab:msmarco_results}) demonstrates BoR applied to Recall@1000, confirming the framework's applicability across success rules.

\section{Conclusion}

Bits-over-Random provides a selectivity metric that reveals when high success rates are misleading. By comparing observed performance to the hypergeometric baseline, BoR exposes the mathematical inevitability: when $K \cdot \bar{R}_q / N$ exceeds 3--5, even perfect retrieval achieves negligible selectivity. Our depth-calibrated identity makes this trade-off explicit and predictable.

For practitioners, we recommend:
\begin{enumerate}
    \item \textbf{Monitor the collapse boundary.} Calculate $\lambda = K \cdot \bar{R}_q / N$ for your system. When $\lambda$ approaches 3--5, you are entering the collapse zone where selectivity becomes mathematically impossible.
    \item \textbf{Use BoR to guide K selection.} Stop increasing $K$ when $\text{BoR}_{\max}$ drops below $\sim$0.1~bits. If $\text{BoR} \approx \text{BoR}_{\max}$, the system has saturated and larger $K$ adds noise, not signal.
    \item \textbf{For tool-based agents: filter aggressively.} With small $N$ (50--500 tools), dumping all definitions into context places the system in the collapse zone. Use two-stage retrieval, dynamic tool loading, or domain-based clustering to keep $\lambda$ small.
    \item \textbf{Report BoR alongside traditional metrics.} High Success@K can coexist with zero selectivity. BoR makes the distinction visible.
\end{enumerate}

Our RAG downstream check demonstrates that achieving high success rates without selectivity proves counterproductive in practice: LLM accuracy degraded by 10--16 percentage points at $K=100$ despite 100\% success, while inflating token costs $\sim$10$\times$.

Future work should extend BoR to multi-evidence requirements ($\geq m$ rules), develop adaptive depth selection under computational budgets, and validate the tool-selection collapse predictions with end-to-end agent benchmarks.

\subsubsection*{Acknowledgments}

We are grateful to Mike Halloran and Himanshu Pathak for their valuable feedback and insightful suggestions.

\bibliography{iclr2026_conference}

@article{brown2012conditional,
  title={Conditional Likelihood Maximisation: A Unifying Framework for Information Theoretic Feature Selection},
  author={Brown, Gavin and Pocock, Adam and Zhao, Ming-Jie and Luján, Mikel},
  journal={Journal of Machine Learning Research},
  volume={13},
  pages={27--66},
  year={2012}
}

@book{manning2008introduction,
  title={Introduction to Information Retrieval},
  author={Manning, Christopher D. and Raghavan, Prabhakar and Schütze, Hinrich},
  year={2008},
  publisher={Cambridge University Press},
  url={https://nlp.stanford.edu/IR-book/information-retrieval-book.html}
}

@inproceedings{formal2022splade,
  title={{SPLADE}: Sparse Lexical and Expansion Model for First Stage Ranking},
  author={Formal, Thibault and Piwowarski, Benjamin and Clinchant, Stéphane},
  booktitle={Proceedings of the 45th International ACM SIGIR Conference on Research and Development in Information Retrieval},
  series={SIGIR '22},
  pages={2288--2292},
  year={2022},
  publisher={ACM},
  address={New York, NY, USA}
}

@inproceedings{lang1995newsweeder,
  title={{NewsWeeder}: Learning to filter netnews},
  author={Lang, Ken},
  booktitle={Proceedings of the 12th International Conference on Machine Learning},
  series={ICML '95},
  pages={331--339},
  year={1995},
  publisher={Morgan Kaufmann},
  address={San Francisco, CA, USA}
}

@article{liu2024lost,
  title={Lost in the Middle: How Language Models Use Long Contexts},
  author={Liu, Nelson F. and Lin, Kevin and Hewitt, John and Paranjape, Ashwin and Bevilacqua, Michele and Petroni, Fabio and Liang, Percy},
  journal={Transactions of the Association for Computational Linguistics},
  volume={12},
  pages={157--173},
  year={2024},
  doi={10.1162/tacl_a_00638}
}

@article{peng2005feature,
  title={Feature Selection Based on Mutual Information: Criteria of Max-Dependency, Max-Relevance, and Min-Redundancy},
  author={Peng, Hanchuan and Long, Fuhui and Ding, Chris H. Q.},
  journal={IEEE Transactions on Pattern Analysis and Machine Intelligence},
  volume={27},
  number={8},
  pages={1226--1238},
  year={2005},
  doi={10.1109/TPAMI.2005.159}
}

@inproceedings{shi2023large,
  title={Large Language Models Can Be Easily Distracted by Irrelevant Context},
  author={Shi, Freda and Chen, Xinyun and Misra, Kanishka and Scales, Nathan and Dohan, David and Chi, Ed H. and Schärli, Nathanael and Zhou, Denny},
  booktitle={International Conference on Machine Learning, ICML 2023, 23-29 July 2023, Honolulu, Hawaii, USA},
  series={Proceedings of Machine Learning Research},
  volume={202},
  pages={31210--31227},
  year={2023},
  publisher={PMLR},
  editor={Krause, Andreas and Brunskill, Emma and Cho, Kyunghyun and Engelhardt, Barbara and Sabato, Sivan and Scarlett, Jonathan},
  url={https://proceedings.mlr.press/v202/shi23a.html}
}

@inproceedings{thakur2021beir,
  title={{BEIR}: A Heterogeneous Benchmark for Zero-shot Evaluation of Information Retrieval Models},
  author={Thakur, Nandan and Reimers, Nils and Rücklé, Andreas and Srivastava, Abhishek and Gurevych, Iryna},
  booktitle={Proceedings of the Neural Information Processing Systems Track on Datasets and Benchmarks 1, NeurIPS Datasets and Benchmarks 2021, December 2021, virtual},
  year={2021},
  editor={Vanschoren, Joaquin and Yeung, Sai-Kit},
  url={https://datasets-benchmarks-proceedings.neurips.cc/paper/2021/hash/65b9eea6e1cc6bb9f0cd2a47751a186f-Abstract-round2.html}
}

@article{truchon2007evaluating,
  title={Evaluating Virtual Screening Methods: Good and Bad Metrics for the Early Recognition Problem},
  author={Truchon, Jean-François and Bayly, Christopher I.},
  journal={Journal of Chemical Information and Modeling},
  volume={47},
  number={2},
  pages={488--508},
  year={2007},
  doi={10.1021/ci600426e},
  url={https://doi.org/10.1021/ci600426e}
}

@article{webber2010similarity,
  title={A similarity measure for indefinite rankings},
  author={Webber, William and Moffat, Alistair and Zobel, Justin},
  journal={ACM Transactions on Information Systems},
  volume={28},
  number={4},
  pages={20:1--20:38},
  year={2010},
  url={https://api.semanticscholar.org/CorpusID:16050561}
}

@misc{anthropic2025toolselection,
  title={Introducing advanced tool use on the Claude Developer Platform},
  author={Anthropic},
  year={2025},
  urldate={Retrieved December 04, 2025},
  url={https://www.anthropic.com/engineering/advanced-tool-use}
}

@inproceedings{nguyen2016ms,                                                                                                  
    title={MS MARCO: A Human Generated MAchine Reading COmprehension Dataset},                                                  
    author={Nguyen, Tri and Rosenberg, Mir and Song, Xia and Gao, Jianfeng and Tiwary, Saurabh and Majumder, Rangan and Deng,   
  Li},                                                                                                                          
    booktitle={Proceedings of the Workshop on Cognitive Computation (CoCo@NIPS)},                                               
    year={2016}                                                                                                                 
  }
\bibliographystyle{iclr2026_conference}

\end{document}